\documentclass[11pt, a4paper]{article}

\usepackage[utf8]{inputenc}
\usepackage{amsmath, amssymb}
\usepackage{amsfonts}
\usepackage{amsthm}
\usepackage[margin= 2.4 cm]{geometry}
\usepackage[colorlinks=true, citecolor=blue, linkcolor=black, urlcolor=black]{hyperref}
\usepackage{graphicx}
\usepackage[dvipsnames]{xcolor}
\usepackage{float}
\usepackage{verbatim}
\usepackage{multirow}
\usepackage{geometry}
\usepackage{subcaption}
\usepackage{enumitem}
\usepackage{booktabs}
\usepackage[round]{natbib}
\usepackage{mathtools}
\mathtoolsset{showonlyrefs}

\theoremstyle{definition}
\newtheorem{theorem}{Theorem}
\newtheorem{lemma}{Lemma}
\newtheorem{corollary}{Corollary}
\newtheorem{proposition}{Proposition}
\theoremstyle{definition}

\DeclareMathOperator*{\sign}{\mathrm{sign}}

\DeclareMathOperator*{\var}{\mathrm{var}}
\DeclareMathOperator*{\pr}{\mathrm{pr}}

\usepackage{ragged2e}

\title{Splitting strategies for post-selection inference}
\author{Daniel G. Rasines\footnote{Corresponding author: daniel.garcia-rasines16@imperial.ac.uk} \and G. Alastair Young}
\date{\textit{Imperial College London}}

\begin{document}

\maketitle

\begin{abstract}
We consider the problem of providing valid inference for a selected parameter in a sparse regression setting. It is well known that classical regression tools can be unreliable in this context due to the bias generated in the selection step. Many approaches have been proposed in recent years to ensure inferential validity. Here, we consider a simple alternative to data splitting based on randomising the response vector, which allows for higher selection and inferential power than the former and is applicable with an arbitrary selection rule. We provide a theoretical and empirical comparison of both methods and derive a Central Limit Theorem for the randomisation approach. Our investigations show that the gain in power can be substantial.
\end{abstract}


\section{Introduction} \label{SEC: intro}

Suppose we have data $Y \sim N(\mu, \sigma^2 I_n)$, where $\mu\in \mathbb{R}^n $, $\sigma^2 > 0$, and $I_n$ is the $n\times n$ identity matrix. We assume that the components of $\mu$ are modelled as a function of $p$ covariates, $\mu_i = g(x_{i1}, \ldots, x_{ip})$ for some unknown $g\colon \mathbb{R}^p\to \mathbb{R}$, and denote by $X = (x_{ij}) \in \mathbb{R}^{n\times p}$ the known, fixed design matrix. In many situations, it is suspected that only a few covariates are truly active, and a preliminary variable-selection step is performed to identify these. Having screened a set of potentially relevant variables, we may want to provide inference for the regression coefficients of the best linear approximation of $\mu$ in the selected model, or some other parameter depending on the output of the selection step. 

If the same data that was used for selection is also used to provide inference for the selected parameter, standard inferential procedures are unreliable, typically leading to overoptimistic results; see e.g. \cite{hong}. Data splitting techniques, whereby a portion of the data is reserved for uncertainty quantification, offer a simple yet effective way to circumvent this problem. Unfortunately, data splitting often leads to procedures with little power both for identifying the active covariates and for providing inference for the selected parameters. In this paper, we study an alternative to data splitting, motivated by the work by \cite{tian}, which provides a more efficient way of splitting the sample information, resulting in more powerful procedures, and which is easy to apply with a general variable-selection method. Our main objective here is to provide a theoretical and empirical comparison between the two information-splitting strategies.

Data splitting is a popular tool in prediction problems, where the hold-out observations are used to assess the accuracy of a predictive model. When the goal is inference rather than prediction, a frequent criticism of data splitting is that different splits can produce different selected models, and therefore two people analysing the same data may end up answering different questions. While this is a valid concern, we stress that selection based on the full data is not free from some level of arbitrariness, as the selection process always involves subjective decisions, including the choice of the selection rule itself, and in many cases a random input independent of the data; see e.g. \cite{wasserman, stability}; and \cite{modelX}. Nevertheless, it is important to keep the effect of the random components low, as failure to do so results in high uncertainty about the relevance of the selected variables. 

The information-splitting technique considered in this paper can be viewed as a variant of data splitting which produces datasets which are more similar to the full sample than those resulting from the latter method, and is therefore potentially less affected by randomness. Furthermore, it entails no extra computational cost with respect to data splitting. The method operates by applying the variable selection algorithm to a randomised version of the data, and then basing inference on the conditional distribution of the data given its randomised form, thereby avoiding any selection bias. The general idea of basing selection on an artificial perturbation of the data in this context was proposed by \cite{tian} as a way of deriving uniformly consistent and powerful inferential procedures, and has become a popular device in the literature. In the original paper, the authors compare the resulting inferential power of randomisation and data splitting in circumstances where it is possible to provide inference conditionally on the selection event, showing the superiority of the former. Here we compare the methods in circumstances where inference discards all the information of the selection split, and is therefore unaffected by the complexity of the selection rule. Concisely, the new procedure works by generating artificial noise $W$ and transforming the data-noise pair $(Y, W)$ into two independent components, both informative about the generative model, so that one is used for selection and the other for inference.

A large number of methods have been proposed in recent years to deal with selection bias. They can be broadly divided into two categories: those which assume that the selection algorithm is of a specific form; and those which provide guarantees for an arbitrary selection rule. An important class of methods in the first group is formed by conditional procedures, as considered above, which are constructed by analysing the conditional distribution of the data given the specified selection event, when this is available. This line of work was started by \cite{lockhartetal} and has subsequently been extended in multiple works such as \cite{leetaylor, loftustaylor, leeetal, fithianetal, tibsetalboots}; and \cite{panigrahietal}.
The second group of methods includes the Post-selection Inference (PoSI) approach of \cite{berketal} and extensions of it \citep{bachoc1, bachoc2}, which achieve uniformly valid inference by maximising over all possible model selection procedures, and are very conservative as a result, as well as the data splitting approach of \cite{rinaldo}, which provides model-free procedures with asymptotically valid guarantees in a random-design setting. \cite{cox_ds} analysed data splitting in a simple inferential problem involving many normal means and found it to be competitive against a natural alternative. Methods based on data splitting have also been considered in more recent works such as \cite{rubin}, \cite{wasserman}, \cite{ignatiadis}, and \cite{DS_DiCiccio}. \cite{fithianetal} observed that inference after data splitting based only on the hold-out observations is inadmissible, being always dominated by data-carving rules, which consider the sampling distribution of the full data, conditional on the selection event. Such data-carving may, however, be very complicated to implement in many situations due to the complexity of the conditional distribution. 

\section{Post-selection inference} \label{SEC: target}

Suppose that, for a given data vector $Y$, a variable-selection algorithm selects a subset $s\subseteq \{1, \ldots, p\}$ of the covariates.
In general, determining an appropriate inferential objective post-selection is not straightforward. Under a linearity assumption, $\mu = X\beta$ for some $\beta = (\beta_1, \ldots, \beta_p)^T\in \mathbb{R}^p$, a natural possibility is to provide inference for the components of $\beta$ associated with the selected variables, $\{\beta_i\colon i\in s\}$. However, this is a difficult problem when $p > n$ as the model is not identifiable. A popular alternative target of inference is the projection parameter, proposed by \cite{berketal}. The projection parameter is the regression parameter of $\mu$ projected onto the subspace spanned by the selected columns of $X$:
\begin{equation}
\beta_s(X) = \arg\min_{z\in \mathbb{R}^{\vert s\vert }}  E \left(\Vert Y - X(s)z  \Vert^2 \right) = \{X(s)^T X(s)\}^{-1} X(s)^T \mu,
\end{equation}
where $X(s)$ is the submatrix of $X$ that contains the selected columns and $\vert s\vert$ denotes the number of selected covariates; it is the best linear predictor of $\mu$ in the selected model. When the model is linear in the selected covariates, $\mu = X(s)\beta_s$ for some $\beta_s\in \mathbb{R}^{\vert s\vert}$, $\beta_s(X)$ is simply $\beta_s$. Otherwise, the interpretation of the projection parameter is less transparent: the $j$-th component of $\beta_s(X)$ may be viewed as the average change of the response when the $j$-th selected covariate increases by one unit, approximated in the selected model. An alternative interpretation is given in \S \ref{SEC: inference_projection} (see also \S 3.2 of \cite{berketal}). When the full model is not linear, one can also consider the projection parameter using the full design matrix, $\beta^F(X) = ( X^T X)^{-1} X^T\mu$ (assuming $X$ has full column rank). Some authors refer to this parameter as the full target, and to the previous one as the partial target. In this case, post-selection inference may be provided for the components of $\beta^F(X)$ associated with the selected variables, $\{\beta^F(X)_i \colon i\in s\}$. Another option, presumably more common in practice, is to proceed under the assumption that $\mu = X(s)\beta_s$ and carry out inference on $\beta_s$. In the random-design case, \cite{rinaldo} develop inferential methods for other choices of the selected parameter which depend on the distribution of the covariates.


Let us denote a generic selected parameter, possibly depending on the design matrix as well as on the selected set, by $h_s(\mu; X)$. Adopting the conditional approach \citep{fithianetal},
we deem an inferential statement about $h_s(\mu; X)$ valid if its error guarantees hold under the conditional distribution of the data given the event that $h_s(\mu; X)$ was selected. For example, a $1 - \alpha$ confidence set $T$ for $h_s(\mu; X)$ is valid if it satisfies
\begin{equation}
\pr\left( h_s(\mu; X) \in T \mid S = s \right) \geq 1 - \alpha,
\end{equation} 
where $S$ is the random set of selected covariates. For simplicity, we will assume that the choice of interest parameter depends only on $s$, so that inference on a given $h_s(\mu; X)$ is required if and only if $S = s$. 

For some popular variable-selection algorithms, such as the lasso or stepwise procedures with fixed tuning parameters, the conditioning event $\{ S = s  \}$ can be studied analytically. Often, it can be written as a union of affine sets; see e.g. \cite{leetaylor, loftustaylor}; and \cite{leeetal}. In most cases, however, this event is too complicated to be explored analytically. Furthermore, even when they can be implemented, conditional methods tend to be very conservative; \cite{length1} show, for instance, that in many cases confidence intervals constructed from the conditional distribution of $Y\mid \{S=s\}$ have infinite expected length. In such cases, data splitting offers an analytically simple and computationally light solution to the inference problem.

\section{Splitting methods} \label{SEC: splitting_methods}

The most common form of data splitting is simple data splitting. Here, a fraction $f = n_1/n$, $1 \leq n_1 < n$, is specified, and a set of indices $R$ is chosen uniformly at random from the subsets of $\{1, \ldots, n\}$ of size $n_1$. Then, for an outcome $R = r$, the sets of observations $(Y^r, X^r)$ and $(Y^{r^c}, X^{r^c})$ are respectively used for selection and for inference, where $Y^r = (Y_i)_{i\in r}$, $X^r = (x_{ij})_{i\in r}$, and $r^c$ denotes the complement of $r$. Since $Y^r$ and $Y^{r^c}$ are independent by assumption, the conditional distribution of the inference set given the output of the selection step is the same as the unconditional one, so classical procedures can be used to provide valid inference for a selected parameter. More elaborated data splitting rules can be found in the prediction literature; see e.g. \cite{data_splitting}. These rules allocate the samples to the selection and inferential sets according to the observed values of the covariates, usually trying to divide them as evenly as possible, to ensure that the analyst has access to similar regions of the design space in both stages. Quite generally, then, a data splitting rule can be formalised as a random variable $R$, possibly depending on $X$, taking values in the power set of $\{1, \ldots, n\}$.

Here we shall consider a different way of distributing the sample information between selection and inference via randomisation. Suppose that $W$ is a random quantity, possibly depending on $X$, and that in the selection step we only allow ourselves to observe the value of a function $U \equiv u(Y, W)$. Since selection depends on the data only through $U$, inference based on the conditional distribution of the data given its observed value, $Y\mid \{U = u\}$, is free of selection bias, and does not require knowledge of the selection mechanism (note that, by contrast, a conditional approach would base inference on $Y\mid \{S(U) = s\}$). 
In particular, we shall be concerned with cases where it is possible to define a quantity $V \equiv v(Y,W)$ which is independent of $U$ and such that $(U, V)$ is sufficient for $Y$, so that inference based on the conditional distribution of $Y\mid \{U = u\}$ is equivalent inference based on the marginal distribution of $V$. 

\cite{tian} proposed randomisation schemes of the form $U = Y  + W$, where $W$ is $n$-dimensional artificial noise whose variance controls the amount of information reserved for inference: small values assign most of the sample information for selection, while large values allocate most of it for inference. One clear advantage of this approach over data splitting is that it gives access to all the observed values of the covariates both at the selection and at the inferential stages, while in data splitting we only have access to a subset of them at each stage. To ensure high inferential power, \cite{tian} recommend that the distribution of $W$ has tails at least as heavy as the normal distribution. If the observation variance $\sigma^2$ is known, a common choice is $W\sim N(0_n, \sigma^2 \gamma I_n)$, where $\gamma > 0$ and $0_n$ is an $n$-dimensional vector of zeroes. This allows for a remarkably simple analysis, as $U\sim N(\mu , \sigma^2  (1 + \gamma)  I_n)$ is of the same parametric form as the data, and basing inference on $Y\mid \{U = u\}$ amounts to basing it on the marginal distribution of $V= Y - \gamma^{-1} W \sim N(\mu ,  \sigma^2 (1 + \gamma^{-1}) I_n)$. This follows because $U$ and $V$ are independent, as they are uncorrelated and normal, and jointly sufficient for $\mu$. 

The previous scheme can of course be generalised by considering an arbitrary normal noise vector $W\sim N(0_n, \sigma^2 \Sigma_W)$, with $\Sigma_W$ positive definite. This produces the split $U = Y + W\sim N(\mu, \sigma^2\{I_n + \Sigma_W\})$ and $V = Y - \Sigma_W^{-1} W \sim N(\mu, \sigma^2\{I_n + \Sigma_W^{-1}\})$. Henceforth, we refer to this randomisation strategy as the $(U, V)$ decomposition. Our goal here is to show that this approach provides a better division of the available information than data splitting.

If $\sigma^2$ is unknown but can be estimated with reasonable precision, an approximate $(U, V)$ decomposition can be achieved by plugging in the variance estimate in the variance of $W$. In \S \ref{SEC: generalised} we consider the asymptotic validity of this approach. In the linear case, if $p$ is small relative to $n$, $\sigma^2$ can be estimated in the classical way. Otherwise we have to resort to high-dimensional alternatives. In our simulation studies we used the estimator implemented in the \texttt{selectiveInference} R package of \cite{selinf_package}, which estimates $\sigma^2$ using the residual sum of squares from a lasso fit with the penalty parameter tuned  by cross-validation. The good performance of this estimator in sparse models was demonstrated in \cite{reid_variance}. Other methods are available; e.g. \cite{varest1} and \cite{varest2}.

\section{Theoretical analysis}

\subsection{Randomisation as information averaging} \label{SEC: averaging}

An appealing feature of the $(U, V)$ decomposition is that it provides a way of averaging information over multiple data splits using a single noise sample, as we show below. This supports the intuition that it provides a more balanced information split than data splitting, and offers a possible a way of selecting the randomisation variance. Moreover, such representation points to a formal advantage in terms of inferential power over the data splits it averages over. In this section we shall extend the discussion to arbitrary regular parametric models for the sake of completeness, with the Gaussian case providing an analytically workable example. 

Let $Y\sim \mathcal{F}(\beta; X)\in \mathbb{R}^n$ be a random vector whose distribution depends on the design $X$ and on a parameter $\beta\in \mathbb{R}^p$; we will only require mild regularity conditions on the model, and in particular we shall not assume that the components of $Y$ are independent. Denote the Fisher information about $\beta$ in $Y$ by $\mathcal{I}_Y(\beta)$. A data split $r$ distributes the total information between the selection and inferential tasks as
\begin{equation}
\mathcal{I}_Y(\beta) = \mathcal{I}_{Y^r}(\beta) + E_{Y^r}[\mathcal{I}_{Y^{r^c}\mid Y^r = y^r}(\beta)] \equiv \mathcal{I}_{r}(\beta) +  \mathcal{I}_{r^c\mid r}(\beta), 
\end{equation}
while a generic randomisation rule $U = u(Y, W)$ divides the information as 
\begin{equation}
\mathcal{I}_Y(\beta) = \mathcal{I}_{U}(\beta) +  E_U[\mathcal{I}_{Y\mid U = u}(\beta)] \equiv \mathcal{I}_{U}(\beta) +  \mathcal{I}_{Y\mid U}(\beta). 
\end{equation}

Consider a collection of data splits $\mathcal{R} = (r_1, \ldots, r_m)$ and a set of positive weights $\mathcal{P} = (p_1, \ldots, p_m)$ adding up to one. We will say that the randomisation rule $U = u(Y, W)$ averages the information over the splits in $\mathcal{R}$ with respect to $\mathcal{P}$ if 
\begin{equation} \label{EQ: average}
\mathcal{I}_{U}(\beta) = \sum_{i=1}^m p_i \mathcal{I}_{r_i}(\beta).
\end{equation}
Note that this also implies that 
\begin{equation} 
\mathcal{I}_{Y\mid U}(\beta) = \sum_{i=1}^m p_i \mathcal{I}_{r_i^c\mid r_i}(\beta). 
\end{equation}

For linear normal models with known covariance the following result can be easily verified. 
\begin{lemma}\label{lemma}
Let $Y\sim N(X\beta, \Sigma)$, where $\Sigma$ is invertible. For a given $(\mathcal{R}, \mathcal{P})$ such that $\cup_{i = 1}^m r_i = \{1, \ldots, n\}$, a randomisation scheme satisfying \eqref{EQ: average} is given by $U = Y + W$ and $V = Y - \Sigma \Sigma_W^{-1}\Sigma^{-1} W $, where $W\sim N(0_n, \Sigma \Sigma_W)$,
\begin{equation}
\Sigma_W = \left\{ \sum_{i = 1}^m p_{i} A_{r_i} \Sigma  \right\}^{-1} - I_n,
\end{equation}
$A_{r_i} = E_{r_i}^T (E_{r_i} \Sigma E_{r_i}^T)^{-1} E_{r_i}$ and $E_{r_i}$ is the $0/1$ matrix such that $Y^{r_i} = E_{r_i} Y$.
\end{lemma}

For the problem considered here, where $\Sigma = \sigma^2 I_n$, we get $\Sigma_W = \Gamma$, where $\Gamma$ is diagonal with $\Gamma_{ii} = w_i^{-1}-1 $ and $w_i = \sum_{i \in r} p_r$. Furthermore, if $\mathcal{R}$ contains all subsets of $\{1, \ldots, n\}$ of size $n_1$ and all the weights are equal, we have that $w_i = n_1/n \equiv f$, so that $\Sigma_W =  (1-f) f^{-1} I_n$.

In the low-dimensional cases, the optimality of the Fisher information is commonly measured through summary statistics of its inverse. In such cases,
an alternative interpretation of \eqref{EQ: average} becomes relevant. Suppose that the sets $\mathcal{R}$, $\mathcal{P}$ represent a random data splitting rule under which $r_i$ is selected with probability $p_i$. Then, any randomisation rule which averages over $\mathcal{R}$ with respect to $\mathcal{P}$ provides a more efficient division of the information than the corresponding random data splitting rule if the optimality of the inverse Fisher information is measured in a particular way. The idea is that, for such $U$, $\mathcal{I}_{ U}(\beta)$ is, on average, more optimal than $\mathcal{I}_{r}(\beta)$, and, similarly, $Y\mid U$ is on average more optimal than $Y^{r^c}\mid Y^r$. 

\begin{proposition} \label{PROP}
Let $R$ be a random data splitting rule induced by $(\mathcal{R}, \mathcal{P})$ and $\varphi$ be a real-valued function defined on the set of $p\times p$ positive definite matrices which is convex and strictly increasing. Let $U = u(Y, W)$ be randomisation scheme that averages over $\mathcal{R}$ with respect to $\mathcal{P}$, and assume that $\mathcal{I}_{r}(\beta)$ and $\mathcal{I}_{r^c\mid r}(\beta)$ are invertible for all $r\in \mathcal{R}$, and that $ \mathcal{I}_{r_1}(\beta) \neq  \mathcal{I}_{r_2}(\beta) $ for some $r_1, r_2\in \mathcal{R}$. Then,
\begin{equation}
\varphi\left\{ \mathcal{I}_{U}(\beta)^{-1}\right\}  <  E\left[ \varphi\left\{ \mathcal{I}_{R}(\beta)^{-1} \right\}\right] \text{    and    }
\varphi\left\{ \mathcal{I}_{Y\mid U}(\beta)^{-1} \right\} <  E\left[\varphi\left\{ \mathcal{I}_{R^c\mid R}(\beta)^{-1} \right\} \right].
\end{equation}
\end{proposition}
 
In the linear Gaussian model, Proposition \ref{PROP} has a direct interpretation in terms of inferential accuracy. Assume that $Y\sim N(X\beta, \sigma^2 I_n)$, with $X^TX$ invertible and $\sigma^2$ known, and, for a given $(\mathcal{R}, \mathcal{P})$, let $U$ and $V$ be defined as in Lemma~\ref{lemma}. Denote by $\hat\beta_{r^c}$ and $\hat\beta_V$ the maximum likelihood estimators of $\beta$ based on $Y^{r^c}$ and $V$, respectively. Note that when providing inference with a data split $r^c$, the estimation variance ought to be considered conditional on the split: $\text{var}(\hat\beta_{R^c}\mid R = r) = \mathcal{I}_{r^c}(\beta)^{-1}$, rather than unconditionally, as $R$ is an ancillary. Taking $\varphi(A) = \eta^T A\eta$ for some $\eta\in\mathbb{R}^p\setminus \{0_n\}$, we get
\begin{equation*}
\text{var}(\eta^T\hat\beta_V) < E[\text{var}(\eta^T\hat\beta_{R^c}\mid R)].
\end{equation*}
In particular, when $\sigma^2$ is known, randomisation produces, on average over the data splits, smaller confidence intervals for any linear combination $\eta^T \beta$ than the data splitting rule it is designed to improve upon. Equal-tailed confidence intervals for $\eta^T\beta$ based on the data split $Y^{r^c}$ are given by $[\eta^T \hat\beta_{r^c} \mp k \mathrm{var}(\eta^T\hat\beta_{r^c})^{1/2} ]$ for some constant $k$, while intervals with the same coverage based on  $V$ are of the form $[\eta^T \hat\beta_V\mp k \mathrm{var}(\eta^T\hat\beta_V)^{1/2} ]$. We can therefore state the result in terms of average confidence interval length.
\begin{corollary}
In the current setting, let $L =  \mathrm{var}(\eta^T\hat\beta_V)$ and $L(r^c) =  \mathrm{var}(\eta^T\hat\beta_{r^c})$. We have that $L < E[L(R^c)]$.
\end{corollary}
Note that this does not say anything about any particular data split $r^c$, which can potentially produce smaller intervals. 

Furthermore, since the maximum likelihood estimators of linear combinations are unbiased, by the Law of Total Variance we also have the unconditional version of the result, where the variance is computed relative the data and the data splitting rule distributions:
\begin{equation*}
 \mathrm{var}(\eta^T\hat\beta_V) < \mathrm{var}(\eta^T\hat\beta_{R^c}).
\end{equation*}
This has a different interpretation: on repeated application of the method, estimates based on $V$ will be, on average, more accurate than estimates based on $Y^{R^c}$.

In other models, an analogous asymptotic interpretation may be given provided the maximum likelihood estimators of $\beta$ based on $U$ and $Y\mid \{U = u\}$ satisfy the Central Limit Theorem. The precise implications of Proposition \ref{PROP} in the selection stage are harder to pinpoint. In \S \ref{SEC: simulation} we conduct  a simulation study to compare data splitting and randomisation in terms of selection power.

\subsection{Data carving}

In \S \ref{SEC: averaging} we compared randomisation and data splitting in situations where all the information contained in the data used for selection is discarded. In contexts where the selection event is known and tractable, it might also be possible to carry out inference in a fully conditional manner. This is, by basing inference on $Y\mid \{S(U) = s\}$ in the case of randomisation and on $Y\mid \{S(Y^r) = s\}$ in the case of data splitting. In the former case the resulting procedures have been shown to avoid the low power characteristic of non-randomised approaches \citep{length2}. When selection is applied to a subset of the observations this is commonly known as data carving \citep{fithianetal}, though for simplicity we will use the term carving to refer to the fully conditional approach based on either type of information split. 

For randomisation one proceeds as follows. Consider the unrestricted mean model $Y\sim N(\mu, \sigma^2 I_n)$, $\mu\in\mathbb{R}^n$, and suppose that inference is sought for $\psi = \eta ^T\mu $ for some $\eta\in \mathbb{R}^n$. Consider a $(U, V)$ decomposition $U = Y + W$, $V = Y -  \Sigma_W^{-1} W$, and for a selection set $s \subseteq\{1, \ldots p\}$ write the selection event as $E = \{u\colon S(u) = s\}$. Let $P_\eta =  \Vert \eta\Vert^{-2} \eta \eta^T$ be the projection matrix onto the line spanned by $\eta$. Carved confidence intervals for $\psi$ can be obtained from the conditional distribution of $\hat\psi = \eta^TY$ given $\{U\in E, (I_n - P_\eta)Y = z\}$, which is free of nuisance parameters. Specifically, let $F_\psi(x) = \pr\{\hat\psi\leq x\mid U\in E, (I_n - P_\eta)Y = z\}$. Then, a confidence interval of coverage $\alpha = q_2 - q_1$ is given by $[a(Y), b(Y)]$, where the endpoints solve $F_{a(Y)}(\hat\psi) = q_1$ and $F_{b(Y)}(\hat\psi) = q_2$. We note that our construction differs slightly from that in \cite{length2} (page 13), which conditions instead on the observed value of $(I_n - P_\eta)U$. A plausible criticism of the latter approach is that the interval is not a function of $Y$ alone and is therefore in violation of the sufficiency principle. 

If the interval for $\psi$ was constructed from the marginal distribution of $V$ alone, via the distribution of $\eta^T V$, it would have length $l(q_1, q_2) = \{\eta^T \Sigma_V \eta\}^{1/2} \{\Phi^{-1}(q_2) - \Phi^{-1}(q_1)\}$, where $\Sigma_V = \sigma^2\{I_n + \Sigma_W^{-1}\}$. Since carved inference incorporates extra information coming from $U\mid \{U\in E\}$, the resulting intervals should intuitively not be larger than $l(q_1, q_2)$. The following result, an extension of Theorem 1 in \cite{length2}, confirms this intuition. 

\begin{proposition}
The confidence interval defined above has $b(Y) - a(Y) \leq l(q_1, q_2)$.
\end{proposition}

For data splitting the matter is more delicate. Assume that selection has been carried out on a subset of the observations $Y^r$, $r\subset \{1, \ldots, n\}$, so that the selection event can be written as $Y^{r}\in E_r$ for some $E_r\subseteq \mathbb{R}^{\vert r \vert}$. Then a fully conditional approach would be based on the distribution $Y\mid \{Y^{r}\in E_r\}$, which involves $n - \vert r\vert$ observations unaffected by selection. This is, however, not enough in general to avoid arbitrarily large confidence intervals. Indeed, define $[a(Y), b(Y)]$ as before with $F_\psi(x) = \pr\{\hat\psi\leq x\mid Y^r\in E_r, (I_n - P_\eta)Y = z\}$. The following proposition follows trivially from Proposition 1 of \cite{length1}. 

\begin{proposition}
Let $E_r\subseteq \mathbb{R}^{\vert r \vert}$ and the selected parameter be $\eta ^T\mu = \eta_r^T \mu_r + \eta_{r^c}^T \mu_{r^c}$ for some $\eta\in \mathbb{R}^n$. For an observed $y$ with $y^r\in E_r$, define $z = (I_n - P_\eta )y$. If $ \inf\{w\in \mathbb{R} \colon z^r + w \eta^r \in E_r\} > -\infty$ or $\sup\{w\in \mathbb{R} \colon z^r + w \eta^r  \in E_r\} < \infty$, then $E[b(Y) - a(Y)] = \infty$. 
\end{proposition}

A point of confusion might arise from the fact that carved intervals in the random sample setting, where $n$ independent copies of $N(\mu, \sigma^2 I)$ are available and use a subset of them for selection, do in fact have finite expected length by the analysis of \cite{length2}. In the regression setting, however, each coordinate $Y_i$ is only informative about its mean $\mu_i$, so the only information about $\mu^{r}$ available in the conditional distribution $Y\mid Y^{r}\in E_r$ comes from a truncated Gaussian and the resulting intervals can be arbitrarily large.

\section{Asymptotic validity of the $(U, V)$ decomposition} \label{SEC: generalised}

In Gaussian models, efficient information splits are easily achievable via an additive perturbation of the data. However, when the observation error and the randomisation noise are not normal, the distributions of $U$ and $Y \mid U$ are generally not available in closed form, complicating the selection and inferential analyses. Furthermore, when the observation variance has to be estimated or the normality assumption is mildly violated, basing inference on the marginal distribution of $V$, as described above, is not formally justified. In this section we provide a set of conditions under which the $(U, V)$ decomposition is asymptotically valid in more general settings.

Suppose that the model is $Y = \mu + \varepsilon$, $\varepsilon = (\varepsilon_1, \ldots, \varepsilon_n)^T$, where the errors are a random sample from an unknown distribution. We assume that the parameter of interest can be written as $\eta^T \mu$ for some vector $\eta\in \mathbb{R}^n$. If the errors are $N(0, \sigma^2)$ and we know $\sigma^2$, exact post-selection inference for $\eta^T \mu$ after selection based on $U = Y + W$, $W \sim N(0_n, \sigma^2 \gamma I_n)$, is provided via $\eta^TV \sim N(\eta^T \mu, \sigma^2 (1 + \gamma^{-1})  \Vert \eta \Vert^2)$, where $\Vert \cdot \Vert$ denotes the Euclidean norm. For simplicity we shall assume that the randomisation variance is of the form $\Sigma_W = \gamma I_n$. Theorem \ref{TH: Generalised_UV} gives conditions under which this approach is asymptotically valid when the errors are not normal and the observation variance is unknown but can be estimated with enough precision. Unless otherwise stated, all the elements involved in the analysis depend on the sample size $n$. Also, for a vector or matrix $A$, $\max(A)$ denotes the maximum absolute entry of $A$, and, for a square matrix $A$, $\lambda_{\min}(A)$ and $\lambda_{\max}(A)$ denote the smallest and largest eigenvalues of $A$.

\begin{theorem}\label{TH: Generalised_UV}
Let $Y = \mu + \varepsilon$, where $\mu\in\mathbb{R}^n$ and the components of $\varepsilon = (\varepsilon_1, \ldots, \varepsilon_n)^T$ are independent and identically distributed with mean zero, variance $\sigma^2$, and $E(\vert \varepsilon_1 \vert^3) < \infty$. Define $U = Y +  W$ and $V = Y - \gamma^{-1}W$, where $W = \hat\sigma  Z$,  $Z\sim N(0_n, \gamma I_n)$ independent of the data, and $\hat \sigma$ is an estimator of $\sigma$ depending only on the first $[n/2]$ observations. Assume that the selection event $\{u\colon S(u) = s\}$ can be written as $\{M^Tu\in \mathcal{E}\}$, where $M$ is an $m\times n$ matrix and $\mathcal{E}\subseteq \mathbb{R}^m$ is convex. Write $M^T = [M_1^T M_2^T]$, where $M_1$ contains the first $[n/2]$ rows of $M$, and assume that for $A \in \{M_1, M_2\}$, $A^TA$ is invertible for all $n$, $\lambda_{\max}\{A^TA\} = O(n)$, $\lambda_{\max}\{(A^TA)^{-1}\} = O(n^{-1})$, $\max(A) = O(1)$, and that $\max(\eta) \Vert \eta \Vert^{-1} = O(n^{-1/2})$. 
If, as $n\to\infty$, $E(\vert \hat\sigma^2 - \sigma^2 \vert ) =  O(n^{-1/2})$ and $\pr(S = s)^{-1} = o(m^{-3/2} n^{1/2} )$, then
\begin{equation}
(1 + \gamma^{-2})^{-1/2} \hat\sigma^{-1}  \Vert \eta \Vert^{-1}  (\eta^TV - \eta^T\mu) \mid \{S = s\} \xrightarrow{d} N(0, 1), \quad n\to \infty.
\end{equation} 
When $\varepsilon_i\sim N(0, \sigma^2)$, the convexity requirement is not needed and the asymptotic requirement on the selection probability can be relaxed to $\pr(S = s)^{-1} = o(m^{-1/2} n^{1/2} )$.
\end{theorem}

Importantly, the requirements on $\hat\sigma$ are independent of the selection events: the assumed lower bound asymptotic  condition on the selection probability ensures that $\hat\sigma$ is consistent for $\sigma$ also conditionally on selection. The reason for estimating $\sigma$ using using only a subset of the observations is to limit the dependence between $\hat\sigma$ and $M^TY$, ensuring that the distribution of the latter is asymptotically Gaussian.
For some standard selection rules such as the lasso, lars, or stepwise regression with fixed hyperparameters, the selection event can be represented in the form described above with $M = X$, in which case the conditions on $M$ are standard. Furthermore, for these rules selection events can be written as convex polytopes after conditioning on the sign of the selected coefficients.
The asymptotic condition on $\eta$ is very natural and ensures that the asymptotic support of $\eta/\Vert \eta\Vert $ is unbounded. When inference is sought for a projection parameter, this condition is satisfied under very mild conditions; see Appendix.

Recently there have been two important proposals for controlling type I error in variable selection: stability selection \citep{stability, stability2} and the fixed-$X$ knockoffs \citep{knockoff}. These methods are not themselves variable-selection algorithms, but modes of implementation of existing ones that ensure error control. Thus far the resulting selection algorithms have been deemed outside the reach of analytical conditional methods due to the untraceability of their selection events. Simulation approaches have been proposed to bypass this problem, but they can be very computationally expensive \citep{blackbox}. In the following proposition we show that, when applied in conjunction with the lasso, the selection events of the respective algorithms can be written in the form required by Theorem 1 after some appropriate conditioning. We refer the reader to the respective articles for details about the algorithms.

\begin{proposition}
Consider the selection functions $S(y; X)$ for the stability selection and knockoff algorithms paired with the lasso. For all $s\subseteq \{1, \ldots, n\}$:
\begin{enumerate}
\item (Stability selection) Suppose that the lasso is applied with a fixed penalty $\lambda > 0$, and fix the set of splits $I_1, \ldots, I_B\subseteq \{1, \ldots, n\}$ to which the algorithm is applied. The events $\{y\colon S(y; X) = s\}\cap \{M^b(y) = m^b\colon b = 1, \ldots, B\}$ are convex for all $\{m^b \colon b= 1, \ldots, B\}$, where $M^b(y) = \sign \{\hat\beta^b(\lambda)\}$ and $\hat\beta^b(\lambda)$ is the lasso solution for the data $[y_{I_b}, X_{I_b}]$. Furthermore, $S(y; X)$ is a function of $My$, with $M = [A_{1} X \ldots, A_B X]$, where $A_i$ is a diagonal matrix with $(A_i)_{jj} = 1$ if $j\in I_{i}$ and $(A_i)_{jj} = 0$ otherwise, and if $\max(X) = O(1)$ and
\begin{equation}
 \min_{i\neq j} \lambda_{\min} (X_{I_i\cap I_j^c}^TX_{I_i\cap I_j^c}) \geq c n,
\end{equation}
for some constant $c$, then $\max(M) = O(1)$ and $\lambda_{\max}\{(M^TM)^{-1}\} = O(n^{-1})$. In this case, $m = B p$.
\item (Knockoffs) The events $\{S(y; X) = s\}\cap \{A(y) = A, s_A(y) = s_A\}$ are convex for all $[A, s_A]$, where $A(y)$ and $s_A(y)$ are the image of the active set of the lasso solution and the image of the set of active signs as $\lambda$ goes from $\infty$ to 0. Furthermore, $S(y; X)$ is a function of $My$, with $M = [X \tilde X]$, where $\tilde X$ is the knockoff copy of $X$, and if $\max(X) = O(1)$, $\lambda_{\max}(X^TX) = O(n)$, and $s_i \leq 2\lambda_{\max}(X^TX) - c n$ for all $i$ and some universal constant $c > 0$, where $s_i$ is the $i$-th diagonal entry of $X^TX - X^T \tilde X$, then $\max(M) = O(1)$ and $\lambda_{\max}\{(M^TM)^{-1}\} = O(n^{-1})$. In this case, $m = 2p$.
\end{enumerate}
\end{proposition}

\section{Simulation study} \label{SEC: simulation}

\subsection{Setting}

We compared data splitting with the $(U, V)$ decomposition in the context of the normal linear model. The data generating process was $Y= X \beta + \epsilon$, with $\epsilon$ distributed as $N(0_n, I_n)$ and  $\beta\in \mathbb{R}^p$ an unknown sparse vector of coefficients. At each replication of the simulations, the rows of the design matrix were generated as independent samples from the distribution $N(0_p, \Gamma)$, where $\Gamma \in \mathbb{R}^{p\times p}$ is a Toeplitz matrix with $(i, j)$ entry $\rho^{\vert i - j \vert}$ for some $\rho\geq 0$. The observation variance, fixed at $1$, was assumed unknown, and was estimated in the classical way when $p < n/4$, and using the high-dimensional alternative otherwise, as per \S \ref{SEC: splitting_methods}. We compared both information splitting strategies in terms of selection power, selection stability, and inferential power in low and high-dimensional settings. All the simulation results can be found in the Appendix.

The data splitting rule considered was the DUPLEX \citep{Snee}. In the even case ($\vert r \vert = n/2$), DUPLEX finds the two covariate observations that are farthest apart and assigns them to the selection set. Then, it finds the next two observations that are farthest apart and assigns them to the inference set. Finally, it allocates the remaining samples one at a time, rotating between the selection and inference sets, by selecting the remaining sample that is farthest apart from the rest of the observations in the given set, starting with the selection set. If the selection and inferential sets have unequal sizes, the algorithm is applied until the smallest set is filled, and all the remaining observations are assigned to the other one. 
For a given splitting fraction $f = \vert r\vert /n$, we compared the performance of DUPLEX with the performance of the $(U, V)$ decomposition with $\Sigma_W = (1-f)f^{-1} I_n$. Recall that such randomisation strategy can be interpreted as a form of averaging information over data splits of the same size $\vert r\vert$ as the data split (\S~\ref{SEC: averaging}).

Thus, for a given $f$ and dataset $(Y, X)$, we considered two procedures. The first bases selection on $(Y^{r}, X^{r})$ and inference on $(Y^{r^c}, X^{r^c})$,
and the second one bases selection on $(U, X) = (Y +  W, X)$ and inference on $(V, X) = (Y - \gamma^{-1} W, X)$, with $W = \hat{\sigma} Z$, where $Z\sim N(0_n, I_n)$ is artificially generated noise independent of the data and $\hat\sigma^2$ is the estimate of model variance.

Regarding selection, we considered the fixed-$X$ knockoff algorithm of \cite{knockoff} and the stability selector paired with the lasso proposed by \cite{stability} and improved by \cite{stability2}, as implemented in the R packages \texttt{knockoff} \citep{knockoff_package} and \texttt{stabs} \citep{stabs_package}. As mentioned earlier, these are selection rules for which conditional inference is analytically intractable and for which computational approaches are very demanding, so they constitute an example where information-splitting approaches would likely be preferred in practice. These algorithms aim at identifying the set of active coefficients while keeping the number of false discoveries under control. The knockoff provides a guarantee on the false discovery rate,
while stability selection controls the expected number of false discoveries. In all the simulations we set the false discovery rate of the knockoff algorithm at $0.3$, and the expected number of false discoveries and cutoff threshold of the stability algorithm at $3$ and $0.7$ respectively. 
Since the knockoff is only applicable in settings with less covariates than observations, it was only considered in the lower dimensional cases.


\subsection{Selection power} \label{SEC: selection_power}

In this simulation we generated $10^3$ triplets $(\beta, Y, X)$ independently for each combination of the following parameters: $n = 200$; $\rho = 0, 0.5$; $f = 1/2, 3/4$ (correspondingly $\gamma = 1, 3^{-1/2}$); $p = 30, 50$ for the knockoff algorithm, and $p = 200, 1000$ for the stability selection algorithm. For each combination of the parameters and each repetition, the true $\beta$ was generated by sampling $10$ non-zero positions uniformly at random and filling them with independent random variables distributed uniformly in the set $\{-1, -0.9, -0.8, \ldots, -0.1, 0.1, \ldots, 0.9, 1\}$. Then, for each $\beta$, a pair $(Y, X)$ was generated, and the corresponding selection algorithm was applied to $(Y^{r}, X^{r})$ in the case of data splitting, and to $(U, X)$ in the case of the randomised procedure. 

The selection ability of the methods was compared according to two criteria: true positive rate and power. The true positive rate is the average number of correct discoveries divided by the total number of active covariates, and the power is defined, for each possible value of $\vert \beta_i\vert = 0.1, \ldots, 1$, as the average number of times a coefficient with absolute value $\vert \beta_i\vert$ is selected, averaged over all the generated coefficients of all $\beta$'s. We also computed, for comparison, the results corresponding to applying the selection algorithms to the full dataset, $(Y, X)$.

\begin{figure}
\centering
\includegraphics[width = \textwidth]{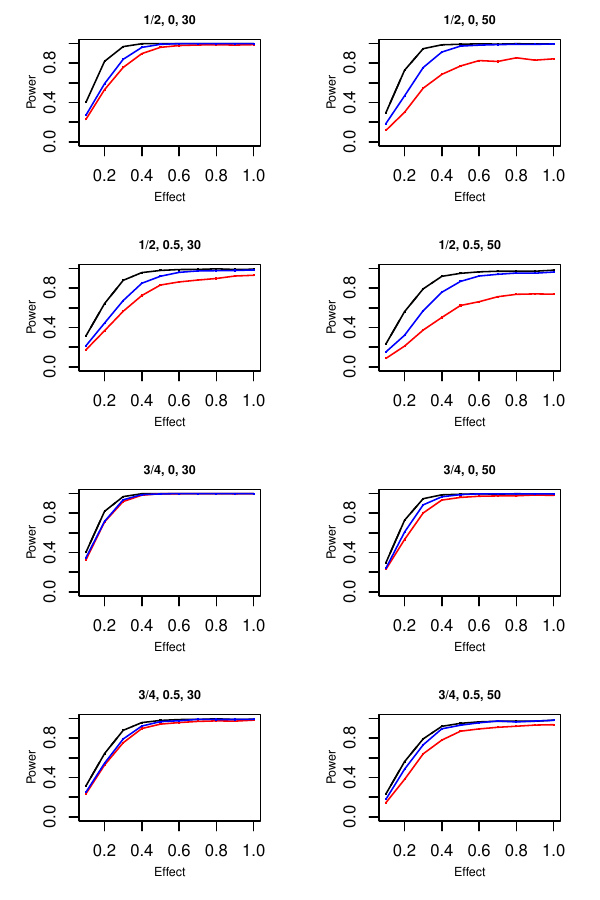}
\caption{Power of the knockoff applied to the complete dataset (black), a data split (red), and a randomised version of the data (blue). The titles indicate $(f, \rho, p)$.}
\label{FIG: power_knockoff}
\end{figure}

\begin{figure}
\centering
\includegraphics[width = \textwidth]{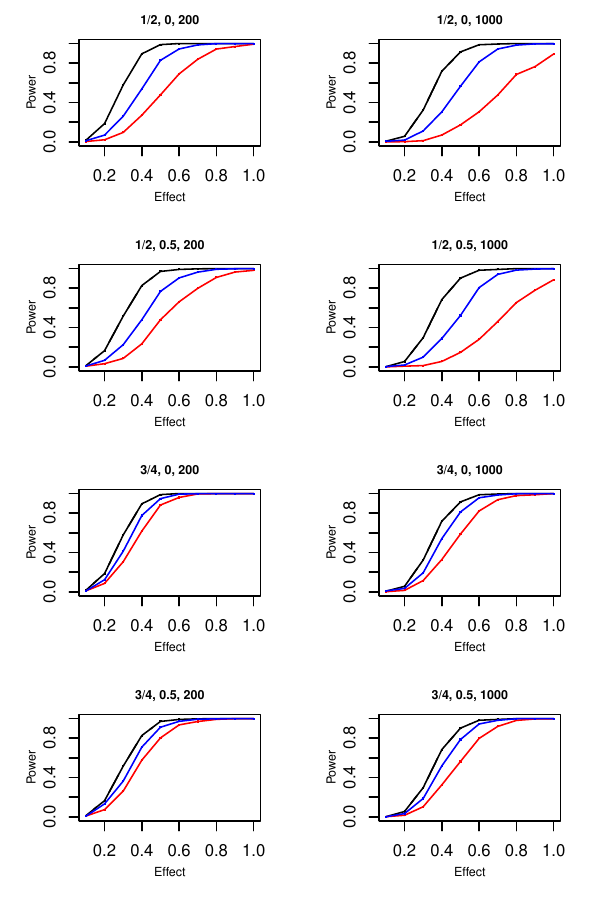}
\caption{Power of stability selection applied to the complete dataset (black), a data split (red), and a randomised version of the data (blue). The titles indicate $(f, \rho, p)$.}
\label{FIG: power_stability}
\end{figure}

Table \ref{TAB: TPR} shows the observed true positive rates of data-splitting and randomisation divided by the observed true positive rates of full-data selection, and Figs. \ref{FIG: power_knockoff} and \ref{FIG: power_stability} show the empirical power functions of the three methods. The results clearly favour the randomisation over data splitting, particularly in the cases with larger values of $p$. Quite remarkably, despite the fact that the choices of $f$ and $\gamma$ were balanced, 
selection based on a randomised split yielded a performance which was in some cases closer to full-data selection than to data splitting selection. 

\subsection{Selection stability} \label{SEC: selection_stability}
    
In addition to enjoying high power, it is important that the selection method does not depend strongly on ancillary components of the analysis: in the case of data splitting, on the choice of the selection and inference sets, and, in the case of randomisation, on the observed value of the artificial noise $W$. To measure the stability of the selection strategies with respect to these elements, a simulation was conducted in which the selection algorithms were applied to multiple splits of the same dataset. Since the DUPLEX is a deterministic allocation rule, in this section we considered a simple data splitting scheme instead.

We set $\beta = (1, 0.9, \ldots, 0.1, 0, \ldots, 0)^T$, $\rho = 0.5$, $f = 1/2, 3/4$, $p = 50$ for the knockoff, and $p = 400$ for stability selection. For each $(f, p)$, we generated $100$ pairs $(Y, X)$ and, for each pair, we sampled $50$ selection sets uniformly at random and $50$ realisations of the randomisation noise $W$, and applied the selection algorithms to each data split and perturbed instance of the data. For each $(Y, X)$, we recorded the average number of times each active covariate was selected across the different splits, an estimate of $\pr(i \in S \vert Y, X)$, $i = 1, \ldots, 10$. The empirical mean and standard deviations of the $100$ estimated averages can be found in Table \ref{TAB: STAB}.

Randomisation was more stable than data splitting, because the corresponding selection probabilities of the active covariates were more concentrated around higher values. For example, for the knockoff with $f = 1/2$, for most values of $(Y, X)$ we had more than a $90\%$ probability of selecting $\beta_3$ under repeated sampling of $W$, as opposed to the $\sim 77\%$ of data splitting.

\subsection{Inference for selected coefficients} \label{SEC: inference_coefficient}

In this simulation we consider the problem of constructing confidence intervals for the selected coefficients $\{\beta_i \colon i\in s\}$. Firstly, we show that a face-value approach, which reports the standard confidence intervals ignoring selection, undercovers coefficients with small effect size $\vert \beta_i \vert $, while the information-splitting techniques provide valid intervals. This gives an illustration of the need to take selection into account in the inferential stage. Secondly, we compare the lengths of the intervals derived from the randomised procedure with respect to those obtained by data splitting. To avoid complications related to confidence interval construction in non-identifiable settings, we shall only consider here low-dimensional cases.

Assume that $X$ has full rank and let $\hat\beta_i = e_i^T (X^TX)^{-1}X^TY$ be the ordinary least squares estimator of $\beta_i$. In a classical, non-selective setting, where the inferential goals are determined prior to the data collection, marginal inference for $\beta_i$ is based on the pivot $T_i = (\hat\beta_i - \beta_i)/\hat\sigma$, which follows a scaled $t$ distribution with $n - p$ degrees of freedom, where $\hat\sigma^2$ is the classical variance estimator. If, however, inference on a given $\beta_i$ is only provided for some data samples, $T_i$ is no longer pivotal, and the resulting confidence intervals can be miscalibrated. To exemplify this, we run the following simulation. We fixed $n = 200$, $p = 30$, true parameter $\beta = (1, -1, 0.5, -0.5, 0.2, -0.2, 0, 0, \ldots, 0)^T$, $\rho = 0, 0.5$, and $f = 1/2, 3/4$, and for each combination of $(\rho, f)$, we generated $5\times 10^3$ pairs $(Y, X)$. We then applied the selection algorithm to each dataset as in the previous sections and constructed classical equal-tailed confidence intervals based on $T_i$ for each selected coefficient, with a nominal coverage of $90\%$.
Table \ref{TAB: COV11} contains the observed coverages of these intervals in the rows indicated by FV (face-value), averaged across coefficients with equal absolute effect. Clearly, the face-value intervals are unreliable when $\vert \beta_i\vert = 0$ or $0.2$, with actual coverages significantly lower than the nominal one in most cases. 

The rows labelled with HD contain the coverages of the intervals constructed using only the hold-out split of the data. In the case of data splitting, the intervals were derived from the $t$-pivots of the hold-out observations, $\hat\beta^{\text{DS}}_i = e_i^T (X^{r^c T}X^{r^c})^{-1}X^{r^c T}Y^{r^c}$. With the current simulation parameters, the number of remaining observations is always larger than the number of covariates, so $X^{r^c}$ has full rank almost surely. For the randomised procedure, we used the approximation $\hat\beta^{\text{R}}_i = e_i^T (X^T X)^{-1}X^TV \dot{\sim} N(\beta_i , (1 + \gamma^{-2}) \sigma^2 e_i^T (X^T X)^{-1} e_i)$. The intervals were thus constructed by studentisation of $\hat\beta^{\text{R}}_i$. Despite the distributional approximation, we see that the coverages of the resulting confidence intervals were very close to the nominal one. 

For the hold-out methods we also recorded the average length of the intervals, which can be found in Table \ref{TAB: LENGTH11}. The intervals provided by the randomised procedure were always shorter, on average, than those provided by the data splitting procedure. In the cases where the amount of information reserved for inference was small ($f = 3/4$), the matrices $X^{r^c}$ were of dimension $50\times 30$, and the resulting intervals were very wide as a consequence. The randomised procedure performed significantly better in these cases, giving intervals roughly $75\%$ shorter than data splitting. The maximum observed standard deviations of the figures shown in the tables were, respectively, $1.9$ and $0.007$.

\subsection{Inference for projection parameters} \label{SEC: inference_projection}

We now consider settings with a number of covariates exceeding the sample size. In these situations the full model is not identifiable, and the methods used in the previous section are not applicable. Instead, we consider the problem of constructing confidence intervals for the coefficients of a projection parameter $\beta_s(X) = \{X(s)^T X(s)\}^{-1} X(s)^T X \beta$.

If $s$ had been fixed in advance and we knew the value of $\sigma^2$, inference on the coefficients of $\beta_s(X)$ would be based on the components of $\hat \beta_s(X) = \{ X(s)^T X(s)\}^{-1} X(s)^T Y$, with distribution, $N(\beta_s(X) , \sigma^2  \{X(s)^T X(s)\}^{-1})$. So, for a nominal coverage of $1-\alpha$, the confidence intervals would be given by $[\hat \beta_s(X)_i \mp  q_{1-\alpha/2}\sigma [e_i^T\{X(s)^T X(s)\}^{-1} e_i]^{1/2}]$, where $ q_{1-\alpha/2}$ is the $1 - \alpha/2$ quantile of a standard normal distribution. When $\sigma^2$ is unknown, we can plug-in the high-dimensional estimate used for the $(U, V)$ decomposition, $\hat{\sigma}^2_{\text{HD}}$, say, and report $[\hat \beta_s(X)_i \mp  q_{1-\alpha/2}\hat\sigma_{\text{HD}} [e_i^T\{X(s)^T X(s)\}^{-1} e_i ]^{1/2}]$. 

When $s$ is data-dependent, hold-out inference can be provided similarly, with the difference that, for data splitting, the inferential target is not the full-projection parameter $\beta_s(X)$, but the projection parameter based on the hold-out observations. Inference is thus provided for $\beta_s(X^{r^c}) = \{ X^{r^c}(s)^T X^{r^c}(s)\}^{-1} X^{r^c}(s)^T \mu_{r^c}$, where $\mu_{r^c} = E(Y^{r^c})$, and is based on $\hat \beta_s^{\text{DS}} (X^{r^c})=  \{ X^{r^c}(s)^T X^{r^c}(s)\}^{-1} X^{r^c}(s)^T Y^{r^c} \sim  N(\beta_s(X^{r^c}) , \sigma^2   \{ X^{r^c}(s)^T X^{r^c}(s)\}^{-1})$. In the case of randomisation, the target parameter is still $\beta_s(X)$, and inference is based on the normal approximation to the distribution of $\hat \beta_s^{\text{R}}(X) = \{X(s)^T X(s)\}^{-1} X(s)^T V \dot\sim  N(\beta_s(X) , (1 + \gamma^{-2})\sigma^2  \{ X(s)^T X(s)\}^{-1})$. In both cases, $\sigma^2$ was approximated by $\hat\sigma^2_{\text{HD}}$.

The simulation parameters were set as in \S \ref{SEC: inference_coefficient} except for the number of covariates, which was $p = 400 $. Here the knockoff was not considered, as it is not applicable with $p > n$. The results can be found in Tables \ref{TAB: COV_PROJ} and \ref{TAB: LENGTH_PROJ}. Note that in Table \ref{TAB: COV_PROJ} the columns indicate the absolute value of the full-model coefficients $\beta_i$, not the coefficients of the projection parameters, which in general depend on $s$. The coverage results were similar to the lower dimensional case, with the difference that the effect of selection was more pronounced here. In one of the cases considered, the coefficients of the projection parameters associated with the null covariates in the full model were almost guaranteed to be missed by the face-value intervals. The hold-out intervals, on the other hand, remained well-calibrated across the different coefficients. Regarding interval lengths, in this case we computed the average of all the intervals produced for a given selection set $s$, and averaged the results over selection sets of equal size in order to establish a more equitable comparison. We see that the average lengths were very similar in this case. Recall, however, that the selection power of the latter method is substantially higher in high-dimensional settings, so in conjunction randomisation dominates data splitting. The maximum standard deviation of the coverage figures were $1.4, 4.5, 1.2$ and $0.4$, for $\vert\beta_i\vert = 0, 0.2, 0.5$, and $1$, respectively. The maximum standard deviation of the length figures was $0.04$. The entries with a dash indicate that no selection set of the corresponding size was selected in the simulation.

\section{Discussion}

Randomisation offers an alternative to data splitting which divides the sample information more efficiently. We have compared randomisation and data splitting with respect to their selection stability, and their selection and inferential power, showing that the dominance of the former can be substantial in settings with a limited amount of information. Our overall conclusion is that inference, as we have analyzed here, based on the marginal distribution of $V$ in the $(U, V)$ decomposition of information offered by randomisation provides a pragmatic and effective approach in the current context. One limitation of randomisation, however, is that implementation requires an effective estimate of the model variance.


Here we have only considered post-selection inference with conditional requirements. It would be interesting to explore how this idea fits within the PoSI framework of \cite{berketal}, where inferential guarantees are unconditional and valid for arbitrary model selection procedures. In this framework the authors construct a ``PoSI constant'' $K$ such that 
\begin{equation*}
\pr(e_j^T\beta_{S} \in [e_j^T\hat\beta_{S}  \mp K ([X_{S}^TX_{S})^{-1}]_{jj}^{1/2}   \hat\sigma] \text{ } \forall \text{ } j=1, \ldots, \vert S\vert )\geq 1 - \alpha
\end{equation*}
for all variable-selection functions $S$, where $\hat\sigma$ is an estimator of $\sigma$ satisfying certain conditions. In the context discussed here, where $S$ depends on $Y$ only through $U$, it is to be expected that there exists a valid PoSI constant $K(\gamma)$ with analogous guarantees which is decreasing in $\gamma$ and smaller than $K$ for all $\gamma > 0$. A similar idea has been fruitfully considered by \cite{zrnic} with other types of randomisation noise, and further work in this direction would be desirable.


\newpage
\section*{Appendix }
\subsection*{Simulation results}

\begin{table}[h]
\begin{center}
\def~{\hphantom{0}}
\caption{True positive rate of the selection algorithms applied after data splitting (DS) and randomisation (R), normalised by the true positive rate of selection applied to the full dataset}{
{\small
\begin{tabular}{ c  c  c  c  c }
\multicolumn{3}{c}{Knockoff} & \multicolumn{2}{c}{Split} \\
$f$ & $\rho$ & $p$ & DS & R \\
 1/2 & 0 & 30 & 0.903 & 0.942 \\
 1/2 & 0 & 50 & 0.738 & 0.923  \\
 1/2 & 0.5 &30  & 0.820 & 0.914  \\
 1/2 & 0.5 & 50 & 0.648 & 0.890  \\   
 3/4 & 0 & 30 & 0.972 & 0.978   \\
 3/4 & 0 & 50 & 0.935&  0.971 \\
 3/4 & 0.5 & 30 & 0.940 &  0.964   \\
 3/4  & 0.5& 50 & 0.890 & 0.970 \\
\end{tabular}
\qquad
\begin{tabular}{ c  c  c  c  c }
\multicolumn{3}{c}{Stability} & \multicolumn{2}{c}{Split} \\
$f$ & $\rho$ & $p$ & DS & R \\
 1/2 & 0 & 200 & 0.694  & 0.867 \\
 1/2 & 0 & 1000 & 0.486 & 0.821  \\
 1/2 & 0.5 & 200  &0.691 & 0.858  \\
 1/2 & 0.5 & 1000 & 0.478 & 0.820  \\
 3/4 & 0 & 200 & 0.895 & 0.948  \\
 3/4 & 0 & 1000 &  0.826 & 0.933  \\
 3/4 & 0.5 & 200 &  0.886 & 0.948  \\
 3/4  & 0.5& 1000 &  0.826 & 0.934  \\ 
\end{tabular}}}
\label{TAB: TPR}
\end{center}
\end{table}

 \begin{table}[H]
 \begin{center}
\def~{\hphantom{0}}
\caption{Means and standard deviations of the estimated selection probabilities for fixed values of $(Y, X)$}{
{
\small
\begin{tabular}{ c  c  c  c  c  c  c  c  c  c  c  c }
 & & \multicolumn{10}{c}{$\beta_i$} \\
\multicolumn{12}{l}{} \\
$f$ & Split & 1 & 0.9 & 0.8 & 0.7 & 0.6 & 0.5 & 0.4 & 0.3 & 0.2 & 0.1 \\
\multicolumn{12}{l}{} \\
 & & \multicolumn{10}{c}{Knockoff} \\
\multicolumn{12}{l}{} \\
\multirow{4}{*}{1/2} & \multirow{2}{*}{DS}  & 0.81  & 0.76  & 0.77 & 0.74 & 0.71  & 0.67  & 0.62 & 0.53 & 0.38  & 0.15 \\
& & (0.08) & (0.09) & (0.09) & (0.09) & (0.09) & (0.11) & (0.11) & (0.16) & (0.16) &  (0.13) \\
& \multirow{2}{*}{R} & 0.98 & 0.92 & 0.98 & 0.95 & 0.95 & 0.92 & 0.89 & 0.77 & 0.57 & 0.25  \\
& & (0.06) & (0.18) & (0.06) & (0.10) & (0.11) & (0.13) & (0.13) & (0.23) & (0.26) & (0.22) \\
\multicolumn{12}{l}{} \\
\multirow{4}{*}{3/4} &  \multirow{2}{*}{DS} & 0.98 & 0.95 & 0.97 & 0.97 & 0.96 & 0.88 & 0.89 & 0.80 & 0.62 & 0.25 \\
 & & (0.14) & (0.22) & (0.17) & (0.17) & (0.20) & (0.33) & (0.31) & (0.40) & (0.49) & (0.44) \\
 &  \multirow{2}{*}{R} & 0.99 & 0.93 & 0.99 & 0.97 & 0.98 & 0.96 & 0.96 & 0.86 & 0.69 & 0.34 \\
  & & (0.04) & (0.19) & (0.04) & (0.10) & (0.08) & (0.10) & (0.11) & (0.22) & (0.29) & (0.32) \\
  \multicolumn{12}{l}{} \\
 & & \multicolumn{10}{c}{Stability} \\
\multicolumn{12}{l}{} \\
\multirow{4}{*}{1/2} & \multirow{2}{*}{DS}  & 1.00  & 1.00  & 1.00 & 1.00 & 0.99  & 0.95  & 0.87 & 0.58 & 0.15 & 0.01 \\
& & (0.00) & (0.00) & (0.00) & (0.01) & (0.02) & (0.10) & (0.17) & (0.28) & (0.18) &  (0.02) \\
& \multirow{2}{*}{R} & 1.00 & 1.00 & 1.00 & 1.00 & 1.00 & 0.98 & 0.95 & 0.76 & 0.30 & 0.04  \\
& & (0.00) & (0.00) & (0.00) & (0.00) & (0.01) & (0.05) & (0.10) & (0.25) & (0.27) & (0.07) \\
\multicolumn{12}{l}{} \\
\multirow{4}{*}{3/4} &  \multirow{2}{*}{DS} & 1.00 & 1.00 & 1.00 & 1.00 & 1.00 & 1.00 & 0.98 & 0.88 & 0.31 & 0.08 \\
 & & (0.00) & (0.00) & (0.00) & (0.00) & (0.00) & (0.00) & (0.14) & (0.33) & (0.47) & (0.27) \\
 &  \multirow{2}{*}{R} & 1.00 & 1.00 & 1.00 & 1.00 & 1.00 & 1.00 & 1.00 & 0.90 & 0.44 & 0.06 \\
  & & (0.00) & (0.00) & (0.00) & (0.00) & (0.00) & (0.01) & (0.02) & (0.20) & (0.37) & (0.14) \\
\end{tabular}}}
\label{TAB: STAB}
\end{center}
\end{table}

\begin{table}
\begin{center}
\def~{\hphantom{0}}
\caption{Coverages of confidence intervals for the selected coefficients}{
{\small
\begin{tabular}{ c  c  c  c  c  c  c  c c c c c}
\multicolumn{8}{l}{} \\
& & & & \multicolumn{8}{c}{$\vert \beta_i\vert$} \\
\multicolumn{12}{l}{} \\
& & & & \multicolumn{4}{c}{Knockoff} & \multicolumn{4}{c}{Stability} \\
\multicolumn{12}{l}{} \\
$f$ & $\rho$ & Split & Method & 0 & 0.2 & 0.5 & 1 & 0 & 0.2 & 0.5 & 1 \\
\multicolumn{8}{l}{} \\
\multirow{4}{*}{1/2} &  \multirow{4}{*}{0} & DS & FV & 68.7 & 90.1 & 90.8 & 89.8 & 39.1 & 80.2 & 91.1& 89.8 \\
& & R & FV & 67.3 & 89.9  & 90.6  & 89.8 & 34.9 & 82.6 & 91.1 & 89.8 \\
& & DS & HD & 89.8  & 91.3 & 90.0 & 90.0 & 90.3 & 90.6 & 90.0 & 90.0  \\
& & R & HD & 89.6 & 89.9  & 89.9 & 90.3 & 90.1 & 89.4 & 89.8 & 89.7  \\
\multicolumn{8}{l}{} \\
\multirow{4}{*}{1/2} &  \multirow{4}{*}{0.5} & DS & FV & 73.9 & 85.2  & 91.0 & 89.9 & 61.7 & 81.6 & 89.5 & 90.0 \\
& & R & FV & 72.0  & 84.0  & 91.2  & 89.7 & 55.8 & 80.1 & 89.5 &  89.9 \\
& & DS & HD & 90.0  & 90.8  & 89.9 & 90.2 & 90.6 & 92.5 & 89.9 & 90.1 \\
& & R & HD & 89.8 &89.2 & 90.0 & 90.4 & 90.8 & 89.6 & 90.5 & 89.9 \\
\multicolumn{8}{l}{} \\
\multirow{4}{*}{3/4} &  \multirow{4}{*}{0} & DS & FV & 59.2 & 91.3  & 90.5  & 89.8 & 21.9  & 83.7 & 90.7 & 89.8  \\
& & R & FV & 57.4  & 91.8 & 90.4  & 89.8 & 14.2 & 85.0 & 90.6  & 89.8 \\
& & DS & HD & 90.3  & 89.7  & 89.7  & 89.6 & 90.8 & 90.2 & 89.6 & 89.6  \\
& & R & HD & 89.8  & 90.2  & 90.0  & 89.6 & 88.5 & 90.4 & 89.9 & 90.3 \\
\multicolumn{8}{l}{} \\
\multirow{4}{*}{3/4} &  \multirow{4}{*}{0.5} & DS & FV &  66.4 & 85.8  & 91.5  & 90.0 & 46.9 & 79.9 & 90.3 & 89.9 \\
& & R & FV & 65.0  & 85.2  & 91.4 & 89.9 & 40.0  & 80.8 & 90.6 & 89.9  \\
& & DS & HD & 90.6  & 90.1  & 90.4 & 89.8  & 88.9  & 89.7 & 90.3 & 89.8 \\
& & R & HD & 89.2 & 90.4 & 89.7 & 89.6 & 90.8  & 91.0 & 89.6 & 90.3 
\end{tabular}}}
\label{TAB: COV11}
\end{center}
\end{table}


\begin{table}
\begin{center}
\def~{\hphantom{0}}
\caption{Average length of confidence intervals for the selected coefficients}{
{\small
\begin{tabular}{ c  c  c  c  c  c  c  c c c c c}
\multicolumn{8}{l}{} \\
\multicolumn{8}{l}{} \\
& & & & \multicolumn{8}{c}{$\vert \beta_i\vert$} \\
\multicolumn{8}{l}{} \\
& & & & \multicolumn{4}{c}{Knockoff} & \multicolumn{4}{c}{Stability} \\
\multicolumn{12}{l}{} \\
\multicolumn{4}{l}{} \\
$f$ & $\rho$ & Split & Method & 0 & 0.2 & 0.5 & 1 & 0 & 0.2 & 0.5 & 1 \\
\multicolumn{8}{l}{} \\
 \multirow{2}{*}{1/2} &  \multirow{2}{*}{0} & DS & HD & 0.391 & 0.390 & 0.391 & 0.391 & 0.389 & 0.391  & 0.391  & 0.392 \\
& & R & HD & 0.356  & 0.355  & 0.358  & 0.358 & 0.360  & 0.354  & 0.358 & 0.358  \\
\multicolumn{8}{l}{} \\
 \multirow{2}{*}{1/2} &  \multirow{2}{*}{0.5} & DS & HD &  0.508 & 0.507 & 0.510 & 0.482 & 0.506 & 0.506  & 0.509 & 0.483  \\
& & R &  HD & 0.457  & 0.455  & 0.459 & 0.436 & 0.459  & 0.458 & 0.457 & 0.438  \\
\multicolumn{8}{l}{} \\
 \multirow{2}{*}{3/4} &  \multirow{2}{*}{0} & DS & HD &  0.653 & 0.651  & 0.650 & 0.651 & 0.657  & 0.648 & 0.651  & 0.651 \\
& & R & HD & 0.503 & 0.502  & 0.506  & 0.506 & 0.507 & 0.500 & 0.507 &  0.507 \\
\multicolumn{8}{l}{} \\
 \multirow{2}{*}{3/4} &  \multirow{2}{*}{0.5} & DS & HD &  0.899 & 0.904 & 0.903 & 0.846 & 0.890 & 0.904 & 0.902 & 0.847 \\
& & R & HD & 0.644 & 0.645  & 0.650 & 0.617 & 0.648 & 0.648  & 0.646 & 0.620 
\end{tabular}}}
\label{TAB: LENGTH11}
\end{center}
\end{table}


\begin{table}
\centering
\def~{\hphantom{0}}
\caption{Coverages of confidence intervals for the coefficients of the projection parameters; stability selection}{
\scalebox{0.9}{
{\small
\begin{tabular}{ c  c  c  c  c  c  c  c }
\multicolumn{8}{l}{} \\
\multicolumn{8}{l}{} \\
& & & & \multicolumn{4}{c}{$\vert \beta_i\vert$} \\
\multicolumn{8}{l}{} \\
$f$ & $\rho$ & Split & Method & 0 & 0.2 & 0.5 & 1 \\
\multicolumn{8}{l}{} \\
\multirow{4}{*}{1/2} &  \multirow{4}{*}{0} & DS & FV & 38.6 & 70.9 & 91.2 & 89.4 \\
& & R & FV & 25.0 & 71.6  & 91.1  & 89.2 \\
& & DS & HD & 89.7  & 86.1 & 89.4 & 90.0  \\
& & R & HD & 87.9 & 89.5  & 89.8 & 89.9  \\
\multicolumn{8}{l}{} \\
\multirow{4}{*}{1/2} &  \multirow{4}{*}{0.5} & DS & FV & 41.7 & 75.0  & 83.9 & 90.6 \\
& & R & FV & 32.5  & 75.3  & 87.5  & 90.8 \\
& & DS & HD & 91.2  & 88.0  & 88.0 & 90.4 \\
& & R & HD & 88.4 & 88.9 & 89.2 & 90.2 \\
\multicolumn{8}{l}{} \\
\multirow{4}{*}{3/4} &  \multirow{4}{*}{0} & DS & FV & 11.7 & 71.2  & 90.9  & 89.1  \\
& & R & FV & 5.5  & 72.8 & 90.3  & 89.1 \\
& & DS & HD & 88.2  & 89.8  & 90.0  & 90.0  \\
& & R & HD & 89.4  & 90.3  & 90.0  & 89.9 \\
\multicolumn{8}{l}{} \\
\multirow{4}{*}{3/4} &  \multirow{4}{*}{0.5} & DS & FV & 18.3 & 71.3  & 88.5 &  90.5 \\
& & R & FV & 12.9 & 70.1 & 89.9 & 90.1  \\
& & DS & HD & 90.0 & 92.3  & 91.4  & 90.5   \\
& & R & HD & 89.0 & 91.3 & 90.4 & 90.3 \\
\end{tabular}}}}
\label{TAB: COV_PROJ}
\end{table}

\begin{table}
\centering
\def~{\hphantom{0}}
\caption{Average length of confidence intervals for the coefficients of the projection parameters; stability selection}{
{\small
\begin{tabular}{ c  c  c  c  c  c  c  c  c  c  c  c }
\multicolumn{12}{l}{} \\
& & & & \multicolumn{8}{c}{$\vert s \vert$} \\
\multicolumn{12}{l}{} \\
$f$ & $\rho$ & Split & Method &  1 & 2 & 3 & 4 & 5 & 6 & 7 & 8 \\
\multicolumn{12}{l}{} \\
 \multirow{2}{*}{1/2} &  \multirow{2}{*}{0} & DS & HD & - & 0.34 & 0.34 & 0.34 & 0.33 & 0.33 & 0.34 & - \\ 
& & R & HD & - & 0.35 & 0.34 & 0.33 & 0.33 & 0.33 & 0.33 & 0.30 \\ 
\multicolumn{12}{l}{} \\
 \multirow{2}{*}{1/2} &  \multirow{2}{*}{0.5} & DS & HD & 0.34 & 0.38 & 0.37 & 0.37 & 0.36 & - & - & - \\
& & R &  HD & 0.35 & 0.39 & 0.37 & 0.36 & 0.36 & 0.35 & 0.34 & - \\ 
\multicolumn{12}{l}{} \\
 \multirow{2}{*}{3/4} &  \multirow{2}{*}{0} & DS & HD & - & 0.48 & 0.48 & 0.48 & 0.48 &0.47 &0.47 &0.48 \\ 
& & R & HD & - & 0.52 & 0.49 & 0.47 & 0.47 & 0.46 & 0.46 & 0.45 \\ 
\multicolumn{12}{l}{} \\
 \multirow{2}{*}{3/4} &  \multirow{2}{*}{0.5} & DS & HD & 0.48 & 0.55 & 0.54 & 0.52 & 0.52 & 0.53 & - & - \\ 
& & R & HD & 0.49& 0.56& 0.54& 0.52& 0.51& 0.48& 0.49 & - \\ 
\end{tabular}}
}
\label{TAB: LENGTH_PROJ}
\end{table}

\newpage
\subsection*{Proof of Proposition 1}

For two symmetric matrices $A$, $B$ write $A \prec B$ if $B - A$ is positive definite. By the harmonic-arithmetic mean inequality for Hermitian matrices \citep{harmonic}, it follows that 
\begin{equation}
 \mathcal{I}_{U}(\beta)^{-1} = \left\{ \sum_{i = 1}^m  p_i   \mathcal{I}_{r_i}(\beta) \right\}^{-1} \prec  \sum_{i = 1}^m p_i  \mathcal{I}_{r_i}(\beta)^{-1} = E\left\{  \mathcal{I}_{R}(\beta)^{-1} \right\}.
 \end{equation}
 The inequality is strict because the $\mathcal{I}_{r_i}(\beta)$'s are not all equal (by assumption). By the monotonicity of $\varphi$, this implies that
\begin{equation}
\varphi\left\{ \mathcal{I}_{U}(\beta)^{-1} \right\}  < \varphi\left[  E\left\{  \mathcal{I}_{R}(\beta)^{-1} \right\} \right].
\end{equation}
Furthermore, since $\varphi$ is convex, 
\begin{equation}
 \varphi\left[  E\left\{  \mathcal{I}_{R}(\beta)^{-1} \right\} \right] \leq E\left[ \varphi\left\{ \mathcal{I}_{R}(\beta)^{-1} \right\}\right] ,
\end{equation}
which shows the first inequality. The proof for $Y\mid U$ is analogous:
 \begin{equation}
  \mathcal{I}_{Y\mid U}(\beta)^{-1} = \left\{ \sum_{i = 1}^m p_i   \mathcal{I}_{r_i^c\mid r_i}(\beta) \right\}^{-1} <  \sum_{i = 1}^m p_i  \mathcal{I}_{r_i^c\mid r_i}(\beta)^{-1} = E\left\{  \mathcal{I}_{R^c \mid R}(\beta)^{-1} \right\}. 
 \end{equation}


\subsection*{Proof of Proposition 2}

Assume without loss that $\Vert \eta\Vert^{2} = 1$. We have $\hat\psi \mid \{ U\in E, (I_n - P_\eta)Y = z\} \stackrel{d}{=} \hat\psi \mid \{  \hat\psi \eta + z + W \in E\}$, where $\stackrel{d}{=}$ denotes equality in distribution. Write $A = \{ \hat\psi \eta + z + W \in E\} = \{  \hat\psi \Sigma_W^{-1/2} \eta + \varepsilon \in  \Sigma_W^{-1/2}(E - z) \}$, where $\varepsilon = \Sigma_W^{-1/2} W \sim N(0, \sigma^2 I_n)$. Write $T = (I_n - P_{\Sigma_W^{-1/2} \eta}) \varepsilon$, $P_{\Sigma_W^{-1/2} \eta} = (\eta^T \Sigma_W^{-1} \eta)^{-1} \Sigma_W^{-1/2} \eta\eta^T  \Sigma_W^{-1/2}$. The conditioning event is
\begin{equation}
A = \left\{  \left[\hat\psi + \frac{\eta^T \Sigma_W^{-1/2} \varepsilon }{ \eta^T \Sigma_W^{-1} \eta}\right] \Sigma_W^{-1/2} \eta + T \in  \Sigma_W^{-1/2}(E - z)\right \}.
\end{equation}
Let $A(t) = A\cap \{T = t\}$ and define $a(y, t)$, $b(y, t)$ as the solutions of $F_x(\hat\psi\mid A(t)) = q_1, q_2$. Suppose that $a(y, t) < a(y)$ for all $t$. This would imply that $q_1 = F_{a(y, t)}(\hat\psi\mid A(t)) < F_{a(y)}(\hat\psi\mid A(t))$. Taking the expectation  with respect to $T$, we would have that $F_{a(y)}(\hat\psi) > 1-q_1$, a contradiction. By the argument in the opposite direction we conclude that, for all $y$, there exists a $t(y)$ such that $a(y, t(y)) = a(y)$. Now, clearly $b(y, t(y)) > b(y)$, as otherwise $[a(Y), b(Y)]$ would have coverage strictly larger than $q_1 - q_2$. Thus, $b(y) - a(y) \leq b(y, t(y)) - a(y, t(y)) \leq \sup_t b(y, t) - a(y, t) $. Since $\eta^T \Sigma_W^{-1/2} \varepsilon$ is independent of $T$, we can apply Theorem 1 of \cite{length2} with $\tau^2 = \{\eta^T \Sigma_W^{-1} \eta\}^{-1}$ to conclude that $b(y, t) - a(y, t)  \leq l(q_1, q_2)$ for all $t$. This concludes the proof.


\subsection*{Proof of Theorem 1}
Assume without loss of generality that the maximum absolute entry of $\eta$ is $1$ for all $n$.

Define the random vectors $U^* = Y + \sigma Z$, $V^* = Y - \gamma^{-1} \sigma Z$, and $W^* = \sigma Z$. Furthermore, let 
\begin{eqnarray}
\varphi_n(x) &=& \pr\{ M^T U \in \mathcal{E},  (1 + \gamma^{-1})^{-1/2} \sigma^{-1}  \Vert \eta \Vert^{-1}  (\eta^TV - \eta^T\mu) \leq x\}, \\
\varphi^*_n(x) &=& \pr \{M^T U^* \in \mathcal{E},  (1 + \gamma^{-1})^{-1/2} \sigma^{-1}  \Vert \eta \Vert^{-1}  (\eta^TV^* - \eta^T\mu) \leq x\}, \\
\tilde \varphi_n(x) &=&  \pr\left\{  N(M^T\mu,  \sigma^2 (1 + \gamma) M^T M  ) \in \mathcal{E} \right\} \pr\left\{ N(0,  1) \leq x \right\}. 
\end{eqnarray}

By Markov's inequality, for all $t > 0$, 
\begin{equation}
\pr\{\vert \hat\sigma^2 - \sigma\vert^2 \geq t \mid S = s\} \leq \frac{\pr\{\vert \hat\sigma^2 - \sigma^2\vert \geq t\}}{\pr(S = s)}  \leq \frac{1}{t}\frac{E[ \vert \hat\sigma^2 - \sigma^2\vert]}{\pr(S = s)} = O\left( \frac{1}{n^{1/2}\pr(S = s)} \right),
\end{equation}
so $\hat\sigma$ is consistent under the assumed asymptotic regime. Therefore we just need to show that  
\begin{equation}
(1 + \gamma^{-1})^{-1/2} \sigma^{-1}  \Vert \eta \Vert^{-1}  (\eta^TV - \eta^T\mu) \vert \{S = s\} \xrightarrow{d} N(0, 1), \quad n\to \infty,
\end{equation} 
and invoke Slutsky's Theorem to conclude the result. To this end we need to bound $\vert \varphi_n(x) - \tilde \varphi_n(x) \vert$ uniformly in $x\in \mathbb{R}$, as 
\begin{eqnarray}
\frac{\varphi_n(x) }{\varphi_n(\infty) } &=& \pr\{(1 + \gamma^{-1})^{-1/2} \sigma^{-1}  \Vert \eta \Vert^{-1}  (\eta^TV - \eta^T\mu) \leq x \mid S_n = s_n\}; \\ \frac{\tilde\varphi_n(x) }{\tilde \varphi_n(\infty) } &=& \pr\left\{ N(0,  1) \leq x \right\}.  
\end{eqnarray}

In the first part of the proof we bound $\vert \varphi^*_n(x) - \tilde \varphi_n(x) \vert = O(n^{-1/2})$ uniformly in $x\in \mathbb{R}$. When the errors are normally distributed, we trivially have that $ \varphi^*_n(x) = \tilde \varphi_n(x) $ for any $\mathcal{E}$. For the more general case, define the $2n$ vector and the $2n\times (m+1)$ matrix 
\begin{equation}
T = \begin{pmatrix}
\varepsilon \\
\gamma^{-1/2} W^*
\end{pmatrix}, \quad
A = \begin{pmatrix}
M & \eta \\
\gamma^{1/2} M & -  \gamma^{-1/2} \eta  
\end{pmatrix}.
\end{equation}
Note that $A$ has full column rank and that 
\begin{equation}
A^T A = \begin{pmatrix}
(1 + \gamma) M^T M & 0_p \\
0_p^T & (1 + \gamma^{-1} ) \Vert \eta \Vert^2
\end{pmatrix}, \quad A^T T = \begin{pmatrix}
M^T U^* \\
\eta^T V^*
\end{pmatrix} - \begin{pmatrix}
M^T\mu \\
\eta^T\mu 
\end{pmatrix} .
\end{equation}
Define also the $(m+1)$-dimensional vectors $P_i = \sigma^{-1} (A^T A)^{-1/2}A^T e_i \varepsilon_i $ for $i = 1, \ldots, n$, and $P_i = \gamma^{-1/2} \sigma^{-1} (A^T A)^{-1/2}A^T e_i W_i $ for $i = n+1, \ldots, 2n$, where $e_i$ is the $i$-th canonical vector of $\mathbb{R}^{2n}$. Clearly, $E(P_i) = 0_{m+1}$ for all $i = 1, \ldots, 2n$. Furthermore, 
\begin{equation}
\sum_{i = 1}^{2n} P_i = \sigma^{-1} (A^T A)^{-1/2}A^T T \equiv Q, \quad \sum_{i = 1}^{2n}\var( P_i ) = I_{m+1}.
\end{equation}
Consider events of the form $\{M^TU^* \in \mathcal{E}, \eta^T V^* \leq c \}$ for $c\in \mathbb{R}$. These events can be equivalently written as $\{Q \in B_c\}$, where
\begin{equation}
B_c = \left\{ x\in \mathbb{R}^{m+1}\colon \sigma(A^T A)^{1/2}x +  \begin{pmatrix}
M^T\mu \\
\eta^T\mu 
\end{pmatrix} \in \mathcal{E} \times (-\infty, c] \right\}.
\end{equation}
Clearly, since $\mathcal{E}$ is convex, $B_c$ is also convex. Thus, by a version of the multivariate Berry–Esseen theorem for convex sets \citep[Theorem 1.1]{Berry-esseen}, it follows that 
\begin{eqnarray}
& & \vert \pr\{  Q \in B_c \} - \pr\{N(0_{m+1}, I_{m+1}) \in B_c\} \vert \\
&\leq &  \frac{\alpha}{\sigma^3} \left\{  E(\vert \varepsilon_1\vert^3 ) \sum_{i = 1}^n  (e_i^T A (A^T A)^{-1} A^T  e_i)^{3/2}   +  E(\vert W_1\vert^3 )  \sum_{i = n + 1}^{2n} e_i^T A (A^T A)^{-1} A^T  e_i)^{3/2}  \right\} \\
& = &  \tilde \alpha  \sum_{i = 1}^{2n}  (e_i^T A (A^T A)^{-1} A^T  e_i)^{3/2},
\end{eqnarray}
for some $\alpha, \tilde\alpha > 0$ independent of $n$. We can bound the terms of the last sum as follows:
\begin{equation}
e_i^T A (A^T A)^{-1} A^T  e_i \leq \Vert e_i^T A\Vert^2   \lambda_{\max}\left\{ \left( A^T A \right)^{-1}\right\} \leq (m + 1)\max(A)^2   \lambda_{\max}\left\{ \left( A^T A \right)^{-1}\right\} .
\end{equation}
Hence,
\begin{equation}
 \sum_{i = 1}^{2n}  (e_i^T A (A^T A)^{-1} A^T  e_i)^{3/2} \leq 2 n (m + 1)^{3/2} \max(A)^{3}   \lambda_{\max}\left\{\left( A^T A \right)^{-1}\right\}^{3/2} = O(m^{3/2}n^{-1/2})
\end{equation}
by the various asymptotic conditions on $M$ and $\eta$. Recall that we are assuming that the entries of $\eta$ are bounded, and note that the eigenvalues of $A^TA$ are those of $M^TM$ multiplied by $1 + \gamma$, together with $(1 + \gamma^{-1})\Vert \eta\Vert^2$. Taking $c = (1 + \gamma^{-1})^{1/2} \sigma \Vert \eta \Vert x + \eta^T \mu$ gives $\vert \varphi^*_n(x) - \tilde \varphi_n(x) \vert = O(m^{3/2}n^{-1/2})$ uniformly in $x\in \mathbb{R}$. 

Now we use the properties of $\hat\sigma$ to bound $\vert \varphi^*_n(x) - \varphi_n(x) \vert$. Following a natural notation, write
\begin{equation}
\varphi_n(x) = \pr\left\{  \begin{bmatrix}  M_2^T U_2 \\ \eta_2^T V_2  \end{bmatrix} + \begin{bmatrix}  M_1^T U^*_1 \\ \eta_1^T V^*_1  \end{bmatrix}   + L\in   \mathcal{E}(x)   \right\}, \quad L =  (\hat\sigma - \sigma) \begin{bmatrix}  M_1^T \\ - \gamma^{-1} \eta_1^T  \end{bmatrix}  Z_1, 
\end{equation}
where $\mathcal{E}(x) = \mathcal{E} \times (-\infty, \eta^T\mu + (1 + \gamma^{-1})^{1/2} \sigma  \Vert \eta \Vert  x]$. Consider the conditional probability 
\begin{equation}
\varphi_n(x; y_1, z_1) =  \pr\left\{ \begin{bmatrix}  M_2^T U_2 \\ \eta_2^T V_2  \end{bmatrix} +  \begin{bmatrix}  M_1^T U^*_1 \\ \eta_1^T V^*_1  \end{bmatrix}   + l \in  \mathcal{E}(x) \mid y_1, z_1  \right\},
\end{equation}
and let similarly
\begin{equation}
\varphi^*_n(x; y_1, z_1) =  \pr\left\{  \begin{bmatrix}  M_2^T U^*_2 \\ \eta_2^T V^*_2  \end{bmatrix} + \begin{bmatrix}  M_1^T U^*_1 \\ \eta_1^T V^*_1  \end{bmatrix}  \in \mathcal{E}(x) \mid y_1, z_1  \right\}.
\end{equation}
By the same asymptotic arguments as before, conditionally on $y_1$, and in particular on $\hat\sigma$, $\begin{bmatrix}  M_2^T U^*_2 \\ \eta_2^T V^*_2  \end{bmatrix}$ is distributed to error $O(m^{3/2}n^{-1/2})$ over convex sets as a a Gaussian random vector with mean $m = E \begin{bmatrix}  M_2^T U^*_2 \\ \eta_2^T V^*_2  \end{bmatrix} + l$ and covariance
\begin{eqnarray}
C &=& C_1 + (\sigma^2 - \hat\sigma^2) C_2 \\
&\equiv & \sigma^2 \begin{pmatrix}
(1+ \gamma) M_2^T M_2  & 0_p \\
0_p^T & (1+ \gamma^{-1}) \Vert \eta_2 \Vert^2  
\end{pmatrix} + (\hat\sigma^2 - \sigma^2) \begin{pmatrix}
\gamma M_2^T M_2 &M_2^T \eta_2 \\
 \eta_2^T M_2 & \gamma^{-1} \Vert \eta_2 \Vert^2 
\end{pmatrix} 
\end{eqnarray}
if $\lambda_{\max}(C^{-1}) = O(n^{-1})$, and by the requirements on $M_2$, $\eta_2$ this happens for all small enough values of $\vert \hat\sigma^2 - \sigma^2 \vert $, as $\lambda_{\min}(C) \geq \lambda_{\min}(C_1) + (\hat\sigma^2 - \sigma^2) \lambda_{\min}(C_2)$. Let $h> 0$ denote the threshold of $\vert \hat\sigma^2 - \sigma^2 \vert$ for which asymptotic normality holds.

By Theorem 1.1 and Proposition 2.1 of \cite{TV}, we can bound the Total Variation distance between $N(m + l, C)$ and $ \begin{bmatrix}  M_2^T U^*_2 \\ \eta_2^T V^*_2  \end{bmatrix} \sim N(m, C_1)$ by
\begin{equation}
 \vert \hat\sigma^2 - \sigma^2\vert   m^{1/2} \lambda_{\max}(C_1^{-1} C_2) + \{l^T C_1^{-1} l\}^{1/2},
\end{equation}
and $\{l^T C_1^{-1} l\}^{1/2} = \vert \hat\sigma - \sigma \vert \Vert C_1^{-1/2} P_1 z_1 \Vert $, $P_1 = [M_1,  - \gamma^{-1} \eta_1]^T$. We have that 
\begin{equation}
E(\Vert C_1^{-1/2} P_1 Z_1 \Vert ) = E(\Vert  A N(0_{m+1}, \gamma I_{m+1}) \Vert)  \leq k \lambda_{\max} (A) m^{1/2}
\end{equation}
for some $k>0$ and $A = (C_1^{-1/2 }  P_1 P_1^T  C_1^{-1/2})^{1/2}$. From the assumptions it follows that $\lambda_{\max} (A) = O(1)$. Thus, $E( \vert \hat\sigma^2 - \sigma^2\vert   m^{1/2} \lambda_{\max}(C_1^{-1} C_2) + \{l^T C_1^{-1} l\}^{1/2}) = O(m^{1/2}n^{-1/2}) $.

Putting all together, we have that
\begin{eqnarray}
\vert  \varphi_n(x) - \varphi^*_n(x) \vert &=& \left \vert  E[\varphi_n(x; Y_1, Z_1) - \varphi^*_n(x; Y_1, Z_1) ] \right \vert \\
&\leq&  E[\vert \varphi_n(x; Y_1, Z_1) - \varphi^*_n(x; Y_1, Z_1)\vert ] \\
&=&  E[\vert \varphi_n(x; Y_1, Z_1) - \varphi^*_n(x; Y_1, Z_1)\vert \mathbf{1}(\vert \hat\sigma - \sigma\vert \leq  h) ] + O(n^{-1/2}) \\
&=& O(m^{3/2} n^{-1/2}) + O(m^{1/2} n^{-1/2}) + O(n^{-1/2})\\
&=&  O(m^{3/2} n^{-1/2}).
\end{eqnarray}

When the errors are exactly Gaussian, $\begin{bmatrix}  M_2^T U^*_2 \\ \eta_2^T V^*_2  \end{bmatrix}$ is Gaussian and the approximation error can be lowered to $O(m^{1/2} n^{-1/2})$, because the first term in the penultimate line is zero.

\subsection*{Proof of the asymptotic requirement on $\eta$} \label{SEC: eta}

Let $\eta_j(s) = X(s) \{X(s)^TX(s)\}^{-1} e_j$, where $e_j\in \mathbb{R}^p$ is the $j$-th vector of the canonical basis. The $j$-th component of the projection parameter for the selection set $s$ is  is $\eta_j(s)^T \mu$, and the $j$-th component of $\beta$ in the linear model $\mu = X\beta$ is $\eta_j(s)^T \mu$ for $s = \{1, \ldots, p\}$.

On one hand, we have that
\begin{eqnarray} 
\Vert \eta_j(s) \Vert^2 &=& e_j^T  \{X(s)^TX(s)\}^{-1} e_j \\
&\geq & \lambda_{\min}\left[\{X(s)^TX(s)\}^{-1} \right] \\
& = & \lambda_{\max}\{X(s)^TX(s)\}^{-1} \\
& \geq & \lambda_{\max}\{X^TX\}^{-1}.
\end{eqnarray}
Thus, if $\lambda_{\max}(X^TX) = O(n)$, $\Vert \eta_j(s) \Vert^{-1} = O(n^{-1/2})$. By similar arguments and the Cauchy–Schwarz inequality, 
\begin{eqnarray} 
\max\{\eta_j(s)\} &=& \max_{1\leq i \leq n} e_i^T X(s)  \{X(s)^TX(s)\}^{-1} e_j \\
& \leq  & \max_{1\leq i \leq n} \Vert  e_i^T X(s) \Vert \Vert  \{X(s)^TX(s)\}^{-1} e_j \Vert \\
& \leq & \vert s\vert^{1/2} \max(X) \left[ e_j^T   \{X(s)^TX(s)\}^{-2} e_j \right]^{1/2} \\ 
& = & \vert s\vert^{1/2} O(n^{-1}). 
\end{eqnarray}
Putting all together, $\max\{\eta_j(s)\} \Vert \eta_j(s) \Vert^{-1} = O(n^{-1/2})$ if either $\vert s\vert = O(1)$ or $p = O(1)$.

\subsection*{Proof of Proposition 4}

1. Stability selection.

It is clear that selection depends on $y$ only through $[X_{I_1}^Ty_{I_1} \ldots X_{I_B}^T y_{I_B}] = M^T y$, and that $\max(M)\leq \max(X) = O(1)$. Now, $M^TM$ can be written as a $B\times B$ block matrix with $(i, j)$ block equal to $X_{I_i\cap I_j}^T X_{I_i\cap I_j}$, the Gram matrix of the observations that are in $I_i\cap I_j$. Consider the $2\times 2$ block matrix
\begin{eqnarray}
B_{ij} & = & \begin{pmatrix}
X_{I_i}^TX_{I_i} &  X_{I_i\cap I_j}^TX_{I_i\cap I_j} \\
X_{I_i\cap I_j}^TX_{I_i\cap I_j} & X_{I_j}^TX_{I_j} 
\end{pmatrix} \\
& = & \begin{pmatrix}
 X_{I_i\cap I_j}^TX_{I_i\cap I_j}&  X_{I_i\cap I_j}^TX_{I_i\cap I_j} \\
 X_{I_i\cap I_j}^TX_{I_i\cap I_j} &  X_{I_i\cap I_j}^TX_{I_i\cap I_j}
\end{pmatrix} + \begin{pmatrix}
X_{I_i\cap I_j^c}^TX_{I_i\cap I_j^c}&  0 \\
0 & X_{I_i^c\cap I_j}^TX_{I_i^c\cap I_j}
\end{pmatrix}.
\end{eqnarray}
We have that $\lambda_{\min}(B_{ij}) \geq \min\{ \lambda_{\min} (X_{I_i\cap I_j^c}^TX_{I_i\cap I_j^c}), \lambda_{\min} (X_{I_i^c\cap I_j}^TX_{I_i^c\cap I_j}) \}$.  Write $M^TM = 2 B^{-1}(B-1)^{-1} \sum_{i\neq j} \tilde B_{ij} $, where $\tilde B_{ij} $ is equal to $M^T M$ in the blocks $(i, i)$, $(i, j)$, $(j, i)$, and $(j, j)$, and zero otherwise. Computing $v^T M^T M v$ for an arbitrary $v$ with $\Vert v \Vert = 1$ shows that
\begin{equation}
\lambda_{\min}(M^TM) \geq  \min_{i\neq j} \lambda_{\min} (X_{I_i\cap I_j^c}^TX_{I_i\cap I_j^c}).
\end{equation}

For $b = 1, \ldots, B$, let $\hat\beta^b(\lambda)$ be the lasso solution derived from the data $[y_b, X_b]$ and penalty $\lambda$. By \cite{leeetal}, the events $\{y\colon M^b(y) = m^b\}$ are convex polytopes, so $\{y\colon M^b(y) = m^b \text{ } \forall \text{ } b = 1, \ldots, B\}$ is also a convex polytope for any given $m^1, \ldots, m^B$. In stability selection the $j$-th variable is selected if and only if 
\begin{equation}
\sum_{b = 1}^B \vert M^b_j(y) \vert \geq t
\end{equation}
for some fixed threshold $t$, so the selection output is determined solely by $\{M^b(y)\colon b= 1, \ldots, B\}$. 

2. Knockoffs.

Since the knockoffs algorithm depends on $y$ via the solution path of the lasso applied with the extended design matrix $M = [X \tilde X]$, it is clear that selection is a function of $ M^Ty$. By \cite{knockoff} (\S 2.1.1), we have that 
\begin{equation}
    M^TM = \begin{pmatrix}
    X^TX & X^TX - d(s) \\
    X^TX - d(s) & X^TX
    \end{pmatrix},
\end{equation}
where $d(s)$ is the diagonal matrix with diagonal $s$, which is required to satisfy $0\preceq d(s) \preceq 2 X^TX$. From this restriction it follows that $\max(M^TM) = O(1)$, which implies that $\max (M) = O(1)$. Furthermore, using the block-matrix determinant formula and that $an\leq \det(X^TX) \leq bn$ for some constants $a$ and $b$, it is easy to show that the second condition is satisfied if $\lambda_{\max}\{2X^TX - d(s)\} = O(n)$ and $\lambda_{\min}\{2X^TX - d(s)\}^{-1} = O(n^{-1})$. The first condition is trivially satisfied, as $\lambda_{\max}\{2X^TX - d(s)\} \leq \lambda_{\max}\{2X^TX \}$. For the other one, we observe that $\lambda_{\min}\{2X^TX - d(s)\} \geq \lambda_{\min}\{2X^TX \} - \max_{i}\{s_i\} $, so we need $s_i \leq 2\lambda_{\max}(X^TX) - c n$ for all $i$ and some fixed $c > 0$.

Now, by the homotopy path algorithm \citep{efronLARS}, the events $\{A(y) = A, s_A(y) = s_A\}$ are determined by an intersection of linear constraints on $y$, i.e. it is a convex polytope. For the last part we need to show that selection is a function of $[A(y), s_A(y)]$ alone. Note that $[A(y), s_A(y)]$ determines the ordering of $[z_1(y), \ldots, z_p(y), \tilde z_1(y), \ldots, \tilde z_p(y)]$. Thus, it clearly determines $\sign\{w_j(y)\}$ for all $j = 1, \ldots, p$, as well as the ordering of $[\vert w_1(y)\vert, \ldots, \vert w_p(y)\vert]$. In particular, knowledge of $[A(y), s_A(y)]$ allows us to evaluate the inequalities $w_j(y) \leq -\vert w_k(y)\vert $ and $w_j(y) \geq \vert w_k(y)\vert $ for any $j, k = 1, \ldots p$, from which the claim follows.

\newpage
\bibliographystyle{agsm}
\bibliography{references}

\end{document}